\documentclass[intlimits,twoside,a4paper]{article}

\usepackage[utf8]{inputenc}

\usepackage[eqsecnum]{cmpj3}
\issue{2026}{29}{2}{23703}
\doinumber{10.5488/CMP.29.23703}
\title[Effect of weak positional disorder on the miniband structure]%
{
Effect of weak positional disorder on the miniband structure of spherical quantum dot chains}
\author[R. Ya. Leshko]
{ R.~Ya.~Leshko\orcid{0000-0002-9072-164X}\thanks{Email: \email{leshkoroman@gmail.com}.}
}
\address{
 Department of Physics and Information Systems, Drohobych Ivan Franko State Pedagogical University, 3~Stryiska~Str., 82100 Drohobych, Ukraine
}


\date{Received 25 January 2026; revised 4 March 2026; accepted 5 March 2026; published 29 June 2026}

\begin{document}

\maketitle

\begin{abstract}
%	A theoretical framework is developed for the electron miniband structure in one-dimensional 
%	chains of spherical quantum dots subjected to weak positional disorder. 
%	Within the tight-binding approximation combined with the effective-medium approach,
%	the stochastic fluctuations of the inter-dot spacing are mapped onto the renormalization
%	of the key Hamiltonian parameters: the hopping integral $  B  $, the overlap integral $  Q  $,
%	and the on-site energy shift $  M  $. Analytical expressions for these disorder-renormalized 
%	parameters are derived by performing an ensemble average over a narrow Gaussian distribution 
%	of positional deviations ($  \sigma \ll a  $).
%	The resulting generalized dispersion relation shows that weak positional disorder causes a 
%	broadening of the minibands. Specifically, for typical fabrication fluctuations 
%	$\sigma = 0.1\,a  $, the miniband width increases by 8--12\% 
%	(depending on the mean inter-dot distance $a$). 
%	At the same time, the sensitivity of the miniband width to disorder decreases 
%	rapidly with increasing lattice period due to the exponential decay of the electron wave 
%	functions. In the considered weak-disorder regime, the Anderson localization length significantly exceeds 
%	the lattice constant, so the miniband states remain delocalized.

%
 A theoretical framework is developed for the electron miniband structure in one-dimensional chains of spherical quantum dots subjected to weak positional disorder. Within the tight-binding approximation and effective-medium approach, stochastic fluctuations of the inter-dot spacing are mapped onto the renormalization of key Hamiltonian parameters: the hopping integral $B$, overlap integral $Q$, and on-site energy shift M. Analytical expressions for these parameters are derived by performing an ensemble average over a narrow Gaussian distribution of positional deviations ($\sigma \ll a$). The resulting generalized dispersion relation reveals that weak positional disorder broadens the minibands. For typical fabrication fluctuations $\sigma = 0.1\,a$, the miniband width increases by 8--12\% , depending on the mean inter-dot distance $a$. This sensitivity to disorder decreases rapidly with increasing lattice period due to the exponential decay of electron wave functions. In this weak-disorder regime, the Anderson localization length significantly exceeds the lattice constant, meaning states remain delocalized.

%
%\keywords Up to six keywords (\href{https://physh.aps.org/browse}{Physics Subject Headings})
\printkeywords
%
%\pacs {73.21.La, 73.22.Dj, 78.67.Hc}
\end{abstract}

\section{Introduction}

Currently, a considerable number of promising high-performance devices 
(photodetectors, solar cells, transistors, optoelectronic circuits) are based on nanostructures \cite{1,2,3,4,5}. 

Among various nanostructured systems, quantum dot (QD) arrays have attracted tremendous attention 
due to their potential applications in high-performance devices such as QD lasers, third-generation 
solar cells, single photon emitters, and QD memories \cite{6}. These three-dimensionally confined 
semiconductor structures exhibit unique quantum effects, including sharp density of states 
and size-tunable optical properties \cite{6}. QD arrays can be fabricated through various approaches,
including self-assembled growth via Stranski-Krastanov mode in molecular beam epitaxy and metal-organic 
chemical vapor deposition, as well as through colloidal synthesis methods.

Particularly promising are ordered QD arrays with precisely controlled positions, sizes, 
and spatial distribution \cite{7}. Unlike randomly distributed self-assembled QDs, ordered arrays
offer several crucial advantages: 
(i) predictable and reproducible optical and electrical properties; 
(ii) reduced inhomogeneous broadening due to uniform QD sizes; 
(iii) precise site control enabling integration into photonic devices; 
and (iv) the ability to engineer miniband structures through controlled interdot coupling. 
In ordered arrays, QDs can be arranged in one-dimensional (1D), two-dimensional (2D),
or three-dimensional (3D) configurations, each offering distinct electronic and optical characteristics. 
These systems exhibit band-like carrier transport and engineered optical absorption across specific energy ranges.
Those properties are particularly valuable for intermediate band solar cells \cite{8}. 
Furthermore, ordered colloidal quantum dot arrays demonstrate nearly polarization-independent absorption, 
which is an advantage for solar energy harvesting applications where incident sunlight is unpolarized \cite{8}.

It is specifically for ordered QDs that the theory of miniband spectra \cite{9,10,11,12,13,14,15,16,17}
of QD superlattices was developed. The electron miniband structure was calculated using various methods, 
including the plane wave method \cite{9,15,17}, the modified augmented plane wave method \cite{10},
and the tight-binding method \cite{14, 16}. 1D \cite{14, 16}, 2D \cite{9, 10, 13, 17} and 3D \cite{11, 12, 15} 
QD configurations were analyzed here. In the works \cite{9,10,11,12,13,14,15}, ordered arrays of identical 
QDs were considered, typically assuming perfect periodicity and uniformity in size, shape, and composition 
to simplify the miniband calculations. These studies focused on symmetric superlattices where 
the primitive unit cell contains a single type of QD, enabling straightforward application of methods 
such as plane waves or tight-binding approximations to derive the electronic band structure.

By contrast, the work \cite{16} explored a more complex QD chain configuration, 
where the primitive cell of the superlattice incorporates two distinct QDs. 
These QDs differ either in their chemical composition or in size.
The tight-binding method was adapted to account those cases.

However, achieving perfectly ordered QD arrays remains a significant technological challenge in practice \cite{7}.
Real QD arrays are invariably characterized by certain degrees of inhomogeneity 
in both quantum dot sizes and in spatial distribution. 
These imperfections can introduce a weak disorder, which can affect the electronic properties, 
such as broadening of minibands and potential localization effects. 
In view of this, the aim of the study is to elucidate the role of weak disorder in modifying 
the miniband structure in one-dimensional chains of spherical QDs.

\section{Heterosystem model and method of calculation}

\begin{figure}[htb]
	\centerline{\includegraphics[width=1\textwidth]{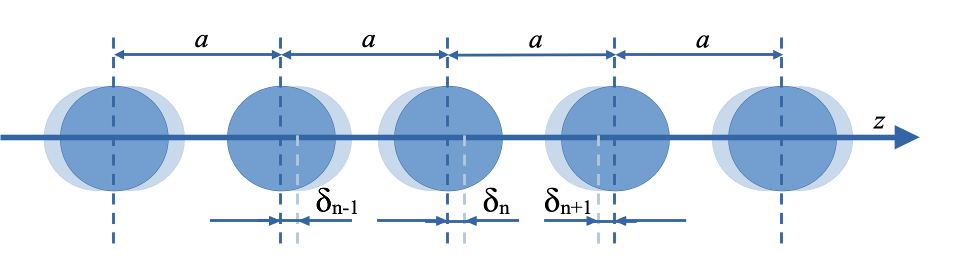}}
	\caption{(Colour online) One-dimensional chains of spherical QDs.} 
	\label{fig1}
\end{figure}

The model describes an infinite periodic chain of spherical QDs with centers at $R_n=a n, n \in \mathbb{Z}$ 
(figure~\ref{fig1}). A weak disorder in the QD positions is taken into account, 
where $\delta_n$ denotes small random deviations from the ideal coordinates $R_n$. 
Each QD has a radius $a_{QD}$. The confinement potential of each QD is chosen as a spherically 
symmetric rectangular potential well. Under the single-level approximation, 
the QD radii are constrained such that each QD possesses only one electronic state.

For an isolated QD (at $R_0 = 0$ ), the Hamiltonian of one QD has form 

\begin{equation}
	\label{H_qd}
	\hat{H}_{QD} = -\frac{\hbar^2}{2}\vec{\nabla}\frac{1}{m(\vec{r})}\vec{\nabla} + U(\vec{r}),
\end{equation}

\noindent where

\begin{equation}
	\label{U_r}
	U(\vec{r}) = \begin{cases}
		-U_0, & \vec{r} \in \text{QD}, \\
		0, & \vec{r} \in \text{matrix},
	\end{cases}
\end{equation}

\noindent is the confinement potential of the QD. 
The electron effective masses within the heterosystem are $m_0$ in the QD and $m_1$ in the matrix, respectively. 
In the isolate QD, electron has energy eigenvalue $E_{QD}$ and the wave function for stationary states

\begin{equation}
	\label{psi_qd}
	\psi_{QD}(\vec{r}) = Y_{00}(\theta, \phi) r^{-1} 
	\begin{cases}
		A_1 {\sin( \xi r)}, & r \leqslant a_{QD}, \\
		A_2 {\re^{-\kappa r}}, & r > a_{QD},
	\end{cases}
\end{equation}

\noindent satisfying the Schr\"{o}dinger equation

\begin{equation}
	\label{schrodinger_qd}
	\hat{H}_{QD} \psi_{QD}(\vec{r}) = E_{QD} \psi_{QD}(\vec{r}),
\end{equation}

\noindent where $Y_{00}(\theta, \phi)$ is the spherical harmonic,

\begin{equation}
	\label{xi_kappa}
	\xi = \sqrt{2 m_0 (U_0 + E_{QD})}/\hbar, \quad \kappa = \sqrt{-2 m_1 E_{QD}}/\hbar.
\end{equation}

The ideal chain consists of QDs at positions $R_n = a n$, where $n = \ldots, -1, 0, 1, \ldots$. 
For the whole QD chain, the electron Hamiltonian has the form:

\begin{equation}
	\label{H_chain}
	\hat{H} = -\frac{\hbar^2}{2}\vec{\nabla}\frac{1}{m(\vec{r})}\vec{\nabla} 
	+ \sum_{n=-\infty}^{\infty} U(\vec{r} - \vec{R}_n).
\end{equation}

Due to translational symmetry, the QD wave functions satisfy the condition $\psi_{QD}(\vec{r} + \vec{R}_n) = \psi_{QD}(\vec{r})$. 
Thus, according to Bloch's theorem, the wavefunction satisfying the Schrödinger equation with
Hamiltonian (\ref{H_chain}) takes the form

\begin{equation}
	\label{psi_chain}
	\Psi_{\vec k}(\vec{r}) = \frac{1}{\sqrt{N}} \sum_{n=-\infty}^{\infty} \re^{\ri \vec{k} \vec{R}_n} \psi_{QD}(\vec{r} - \vec{R}_n),
\end{equation}

\noindent where $\vec{k}$ is the electron wave vector in the first Brillouin zone $  -\piup < \vec{k} \vec{a} < \piup$,
and $N$ is the number of QDs in the chain. 
Substituting this function into the Schrödinger equation, multiplying by the complex conjugate function, 
and using the nearest-neighbors approximation yields the electron energy as a function of the wave vector, 
like in \cite{14,16}:

\begin{equation}
	\label{E_k}
	E(\vec{k}) = E_{QD} + 
	\frac{2 M + 2 B \cos(\vec{k} \vec{a})}{1 + 2 Q \cos(\vec{k} \vec{a})},
\end{equation}

\noindent where the parameters are defined as follows:

\begin{equation}
	\label{M_B_Q}
	\begin{aligned}
		M &= \left\langle \psi_{QD}(\vec{r}) \middle| U(\vec{r} - \vec{a}) \middle| \psi_{QD}(\vec{r}) \right\rangle, \\
		B &= \left\langle \psi_{QD}(\vec{r}) \middle| U(\vec{r} - \vec{a}) \middle| \psi_{QD}(\vec{r} - \vec{a}) \right\rangle, \\
		Q &= \left\langle \psi_{QD}(\vec{r}) \middle| \psi_{QD}(\vec{r} - \vec{a}) \right\rangle.
	\end{aligned}
\end{equation}

The nearest-neighbor approximation remains valid throughout the considered range of inter-dot 
distances ($a = 40-60$ \AA). Numerical estimates show that the overlap integrals with 
next-nearest neighbors are at least one order of magnitude smaller than those with nearest 
neighbors due to the exponential decay of the wave function outside the QD \cite{14, 16}.

Equation (\ref{E_k}) describes the ideal case where all QDs are identical and arranged strictly periodically. 
However, in the presence of weak positional disorder (small random deviations $\delta_n$ from the ideal positions
$R_n$ in $Oz$ direction), the model can be extended. Then, $R_n = a n + \delta_n$.
The deviations $\delta_n$ are assumed to be independent random variables following a normal distribution: 

\begin{equation}
	\label{norma_dist}
	\delta_n \sim \mathcal{N}(0, \sigma^2), \qquad
	P(\delta_n) = \frac{1}{\sqrt{2 \piup} \sigma} \exp\left(-\frac{\delta_n^2}{2 \sigma^2}\right).
\end{equation}

\noindent 
The Gaussian distribution for positional deviations is chosen because it naturally 
describes statistical fluctuations arising during epitaxial growth and colloidal 
synthesis of QD arrays, as confirmed by high-resolution TEM and AFM studies of 
self-assembled QDs \cite{7,19,20}.
Here, $\sigma$ is the standard deviation, quantifying the degree of disorder in the QD positions.
Under the condition of weak disorder ($\sigma \ll a$), the distance $d_n$ between neighboring QDs,
which determines the relevant integrals ($B$, $M$, $Q$), is given by:

\begin{equation}
	\label{d_n}
	d_n = |{R_{n+1} - R_n}| = a + (\delta_{n+1} - \delta_n) = a + \delta,
\end{equation}

\noindent where $\delta \ll a$ too. Since $\delta_{n+1}$ and $\delta_n$ are independent, 
the variance of their difference sums is as follows:

\begin{equation}
	\label{delta}
	\delta = \delta_{n+1} - \delta_n \sim \mathcal{N}(0, 2 \sigma^2). 
\end{equation}

Since the integrals in (\ref{M_B_Q}) depend on the distance between QDs, they can be expanded 
in a series for small deviations. Let $X(d)$ denote a general parameter (where $X$ stands for $Q$, $M$, or $B$), 
then

\begin{equation}
	\label{X_d}
	X( a + \delta ) = X( a ) + X'( a ) \delta + \frac{1}{2} X''( a ) \delta^2 + \frac{1}{3} X'''( a ) \delta^3 + \ldots\, .
\end{equation}

\noindent Averaging over the Gaussian distribution yields $\langle \delta \rangle = 0$, 
$\langle \delta^2 \rangle = 2 \sigma^2$, $\langle \delta^3 \rangle = 0$, etc. Hence,

\begin{equation}
	\label{X_avg}
	\langle X \rangle = X(a) + \frac{1}{2} X''(a) \langle \delta^2 \rangle + O(\sigma^4) = X(a) + X''(a) \sigma^2 + O(\sigma^4).
\end{equation}

The variance of the parameter (to estimate fluctuations) is given by:

\begin{equation}
	\label{var_X}
	\text{Var}(X) = \langle X^2 \rangle - \langle X \rangle^2 = X'(a)^2 \langle \delta^2 \rangle + O(\sigma^4) = 2 X'(a)^2 \sigma^2 + O(\sigma^4).
\end{equation}

In the case of weak disorder, the parameters are replaced by their average values: $\langle M \rangle$, $\langle B \rangle$, and $\langle Q \rangle$. The effective dispersion relation is then given by:

\begin{eqnarray}
	\label{E_k_disorder}
\langle E( k ) \rangle &=& E_0 + \frac{2 \langle M \rangle + 2 \langle B \rangle \cos( k a )}{1 + 2 \langle Q \rangle \cos( k a )}  \nonumber\\
&	\approx& E_0 + \frac{2 [ M(a) + M''(a) \sigma^2 ] + 2 [ B(a) + B''(a) \sigma^2 ] \cos( k a )}{1 + 2 [ Q(a) + Q''(a) \sigma^2 ] \cos( k a )}.
\end{eqnarray}

\noindent Equation (\ref{E_k_disorder}) takes into account that $\vec{a} = (0, 0, a)$. 
Therefore, equation~(\ref{E_k_disorder}) allows for the determination of the electron dispersion 
relation in the chain of QDs in the presence of weak positional fluctuations.

\begin{figure}[!t]
	\centerline{\includegraphics[width=1\textwidth]{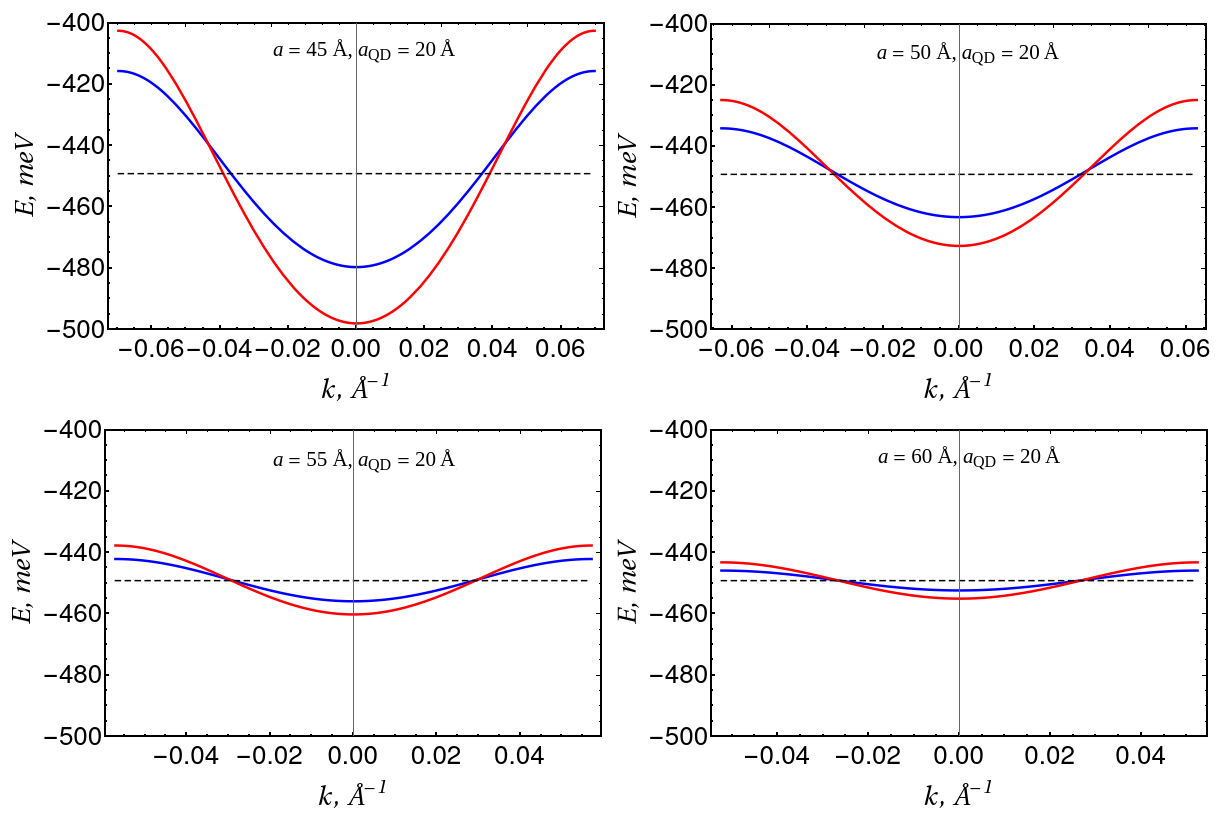}}
	\caption{(Colour online) Electron miniband structure of an ideal (blue curves 1) and a disordered (red curves 2) spherical QD chain.} 
	\label{fig2}
\end{figure}
\begin{figure}[!t]
	\centerline{\includegraphics[width=0.5\textwidth]{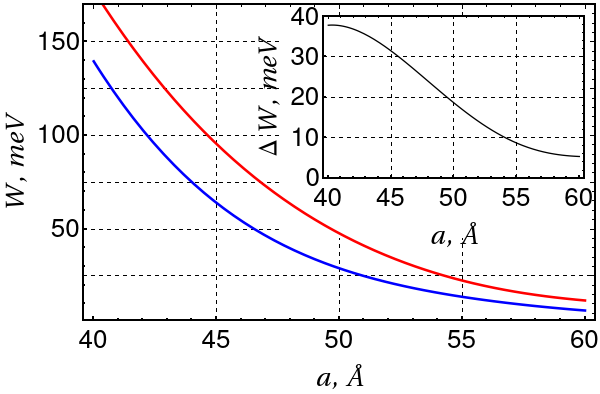}}
	\caption{(Colour online) Electron miniband width calculated with (red curve 1) and without (blue curve~2) disorder. The inset shows the difference in miniband widths between the disordered and ideal cases.} 
	\label{fig3}
\end{figure}
\begin{figure}[!t]
	\centerline{\includegraphics[width=1\textwidth]{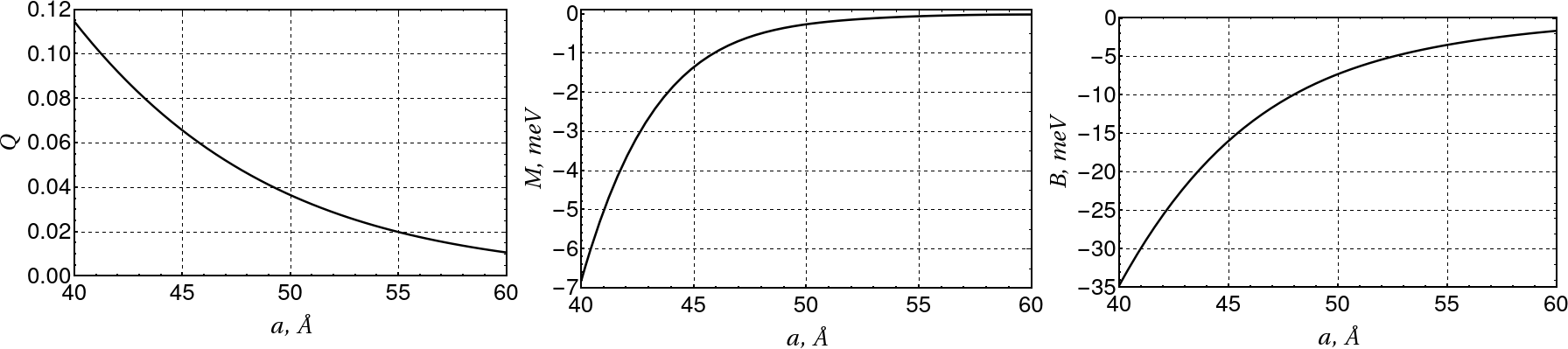}}
	\caption{ Dependence of integrals $Q$, $M$, and $B$ on the period $a$.} 
	\label{fig4}
\end{figure}

\section{Analysis of the obtained results}

All calculations presented in this work were performed at very low temperatures 
($T \to 0$~K). At finite temperatures, phonon scattering and thermal broadening of the 
levels would additionally widen the minibands, potentially masking the weak-disorder 
effects considered (see, for example, \cite{21}).  
%considered, for example, here \cite{21}.

To demonstrate the theoretical results, numerical calculations were performed for a chain of spherical 
GaAs QDs embedded in an AlAs matrix. The following material parameters were used: 
$m_0 = 0.063 m_\text{e}$~\cite{18}, $m_1 = 0.15 m_\text{e}$ \cite{18}, and a potential barrier height of 
$U_0 = 968$ meV \cite{18}, where $m_\text{e}$ is a free electron mass.
 With these parameters, the QD has a single bound electronic level when its radius is less than 21~\AA.
On the other hand, extremely small QD radii limit the applicability of the effective mass approximation.
Therefore, a radius of 20~\AA~was selected for the calculations.

Figure~\ref{fig2} shows the energy minibands of the electron (bands formed by the periodic arrangement of QDs) 
as a function of the wave vector. The effect of possible positional disorder is also included. 
The parameter $\sigma$ in distribution (\ref{delta}) is chosen to be ten times smaller than 
the  distance between QDs ($\sigma = a/10$). Thus, within the considered range of $a$, 
the condition of weak disorder is satisfied.

Figure~\ref{fig2} also demonstrates that disorder induces a broadening of the miniband. 
In the effective-medium approximation, the positional disorder leads to renormalization 
of the average parameters and, consequently, to an apparent broadening of the minibands. 
Importantly, the Anderson localization length in the weak-disorder regime considerably
exceeds the lattice period ($\xi_\text{loc} \gg a$), so the miniband states remain delocalized. 
True Anderson localization would require stronger disorder ($\sigma/a \gtrsim 0.3$), 
which lies beyond the scope of the present study \cite{22}.
Figure~\ref{fig3} displays the miniband width $W = E( \piup/a ) - E( 0 )$ as a function of the parameter 
$a$ for both the disordered system and the ideal QD chain. 
Both figures indicate that an increase in the QD period $a$ results in a decrease in the miniband width. 
Furthermore, the inset in figure~\ref{fig3} highlights the discrepancy between the widths calculated with 
and without taking disorder into account ($\Delta W= W_{\text{disorder}} - W_{\text{ideal}}$). It is evident from both figures that the inclusion of disorder 
leads to a broadening of the miniband compared to the ideal case.
However, as the superlattice period increases (scaling the variance as  to maintain the weak disorder regime), 
the difference between the miniband widths diminishes. This convergence is attributed to the decrease 
in the absolute values of the integrals $Q$, $M$, and $B$ (figure~\ref{fig4}) with increasing inter-dot separation.

The renormalization of the tight-binding parameters due to disorder has a physical interpretation. 
The term proportional to $\sigma^2$ in the expression for the hopping integral $B$ effectively reduces 
the bandwidth. This can be understood as the suppression of coherent tunneling due to the mismatch 
of phases in the wave functions caused by random spacing.
 At the same time, the effective onsite energy shifts due to the modification of the confinement potential 
 by neighboring QDs (terms containing $M$ and $Q$). 
 The interplay of these effects leads to the observed broadening of the minibands.

\section{Conclusion}

This study presents a theoretical approach to model the 
electronic energy spectrum in chains of spherical quantum dots affected by weak positional disorder.
By applying the tight-binding approximation in conjunction with an effective medium scheme, 
we obtained explicit formulas for the adjusted tunneling integral $B$, overlap parameter $Q$, 
and site energy correction $M$ assuming a Gaussian spread in the spacing between dots. 
The calculations reveal that such a disorder leads to a widening of the minibands,
arising from variations in tunneling rates and the overall potential profile,
 yet the band structure remains intact under weak perturbations where the Anderson localization length by
 far surpasses the average lattice spacing. 
 Additionally, it was found that the miniband width becomes less vulnerable to disorder as 
 the inter-dot distance grows, owing to the rapid exponential fall-off of the electron wavefunctions. 
 Overall, these findings offer valuable tools for assessing the manufacturing tolerances and serve 
 as practical recommendations for enhancing the efficiency of quantum dot-based optoelectronic components, 
 including cascade lasers and infrared detectors.

% \bibliographystyle{cmpj}
% \bibliography{cmpjxampl}

%
%% If you have problems with typesetting in ukrainian uncomment lines below.
%
%  \lastpage
%  \end{document}

\ukrainianpart

\title{Вплив слабкого позиційного невпорядкування на мінізонну структуру в одновимірних ланцюжках сферичних квантових точок}
\author{
	Р.~Я.~Лешко
}
\address{
 Кафедра фізики та інформаційних систем, Дрогобицький державний педагогічний університет імені Івана Франка,
вул. Стрийська, 3, 82100 Дрогобич, Україна
}

\makeukrtitle

\begin{abstract}
\tolerance=3000%
  Розроблено теоретичну модель структури електронних мінізон в одновимірних ланцюжках сферичних квантових точок за умов слабкого позиційного безладу. У межах наближення сильного зв’язку та методу ефективного середовища стохастичні флуктуації відстані між точками відображаються на перенормування ключових параметрів гамільтоніана: інтеграла перескоку $B$, інтеграла перекриття $Q$ та зсуву енергії на вузлі $M$. Аналітичні вирази для цих параметрів отримано шляхом усереднення за ансамблем для вузького гауссового розподілу позиційних відхилень ($\sigma \ll a$). Отримане узагальнене дисперсійне співвідношення показує, що слабкий позиційний безлад викликає розширення мінізон. Зокрема, для типових технологічних флуктуацій $\sigma = 0.1a$ ширина мінізони зростає на 8--12\% (залежно від середньої відстані між точками $a$). Водночас чутливість ширини мінізони до безладу швидко зменшується зі збільшенням періоду ґратки через експоненціальне загасання хвильових функцій електрона. У розглянутому режимі слабкого безладу довжина локалізації Андерсона значно перевищує сталу ґратки, тому стани мінізони залишаються делокалізованими.
\keywords неконцентрична сферична квантова точка типу ``ядро-оболонка'', 
електричне поле

\end{abstract}

\end{document}